%% file: QCD-10-012_temp.tex
\pdfoutput=1

\documentclass[11pt,twoside,a4paper,cmspaper,final,collab]{cms-tdr}
\def\svnVersion{71887}\def\cmsCernNo{2011-044}\def\cmsCernDate{\today}\def\cmsMessage{Submitted to Physics Letters B}

\begin{document}\cmsNoteHeader{QCD-10-012}

\hyphenation{had-ron-i-za-tion}
\hyphenation{cal-or-i-me-ter}
\hyphenation{de-vices}
\RCS$Revision: 70591 $
\RCS$HeadURL: svn+ssh://alverson@svn.cern.ch/reps/tdr2/papers/QCD-10-012/trunk/QCD-10-012.tex $
\RCS$Id: QCD-10-012.tex 70591 2011-07-20 13:27:44Z alverson $

\cmsNoteHeader{QCD-10-012} 
\title{Measurement of the Ratio of the 3-jet to 2-jet Cross Sections in pp Collisions at \texorpdfstring{$\sqrt{s}=7$ \TeV}{sqrt(s) = 7 TeV}}

\newcommand {\Ht}{\ensuremath{H_\mathrm{T}} }
\newcommand {\Ratio}{\ensuremath{R_\mathrm{32}} }
\newcommand {\Pt}{\ensuremath{p_\mathrm{T}} }

\date{\today}

\abstract{
A measurement of the ratio of the inclusive 3-jet to 2-jet cross
sections as a function of the total jet transverse momentum,
$H_{\rm T}$, in the range $0.2 < H_{\rm T} < 2.5 {\rm ~\TeV}$
is presented. The data have been collected at a proton-proton centre-of-mass
energy of $7\,\mathrm{\TeV}$ with the CMS detector at the LHC,
and correspond to an integrated luminosity of $36 \ensuremath{\,\text{pb}^{-1}}$.
Comparisons are made between the data and the predictions of different
QCD-based Monte Carlo models for multijet production. All models considered
in this study are consistent with the data for $H_{\rm T} > 0.5 {\rm ~\TeV}$.
This measurement extends to an $H_{\rm T}$ range that has not been
explored before.
}

\hypersetup{%
pdfauthor={CMS Collaboration},%
pdftitle={Measurement of the Ratio of the 3-jet to 2-jet Cross Sections in pp Collisions at sqrt(s) = 7 TeV},%
pdfsubject={CMS},%
pdfkeywords={CMS, physics, QCD, multijets ratio}}

\maketitle 

\section{Introduction}

In leading-order (LO) perturbative Quantum Chromodynamics (pQCD), jet production in pp
collisions occurs when two partons interact via the strong force to
produce two final-state partons. The elementary processes that contribute
in the final state are $\it qq \rightarrow qq$, $\it qg \rightarrow qg$,
$\it gg \rightarrow gg$, and $\it gg \rightarrow qq$,
where $\it q$  and $\it g$ are quarks (or antiquarks) and gluons, respectively.
Each of the final state particles may subsequently lose
energy by emitting other quarks and gluons in a process referred to as a parton
shower (PS). Finally, the products of the parton shower undergo hadronisation
and form hadron jets. Events with three or more jets in the final state
originate from hard-gluon radiation and other higher-order QCD processes.
The jet transverse momentum ($\Pt$), azimuthal angle in the plane perpendicular
to the beams ($\phi$), and rapidity  ($y =  \frac{1}{2} \ln [[E+p_{z}]/ [E-p_{z}]] $,
where $E$ is the jet energy and $p_z$ is the momentum component in the direction
of the colliding protons) are strongly correlated with those of the final-state partons.
For this reason jet measurements have been used to test
pQCD~\cite{bib:UA1bib1, bib:UA2bib1, bib:CDFbib1,  bib:D0bib2, bib:D0, bib:ZEUSbib2,
bib:D0bib1, bib:ZEUSbib1}.

This letter presents a measurement of the inclusive 3-jet to 2-jet cross section ratio,
$R_{32}$, in proton-proton collisions at $\sqrt{s}=7\,\mathrm{\TeV}$.
The measurement is performed using a multijet data sample collected in 2010 by
the Compact Muon Solenoid (CMS) experiment, at the Large Hadron Collider (LHC),
and corresponds to an integrated luminosity of $36\ensuremath{\,\text{pb}^{-1}}$.
Jets are reconstructed in the CMS detector using the anti-$k_T$ jet algorithm~\cite{bib:antikt, bib:fastjet},
with  the requirements $\Pt > 50 {\rm ~\GeV}$ and rapidity  $\vert y \vert < 2.5$.
This cross section definition is chosen to maximise the phase space of the
measurement. The value of $R_{32}$ is measured as a function of the total jet transverse
momentum of the event, $H_{\rm T}$, defined as

\begin{equation}
H_{\rm T} = \sum_{i=1}^{N} {\Pt}_i ~,
\end{equation}

where ${\Pt}_i$ are the transverse momenta of all jets found
with $\Pt > 50 {\rm ~\GeV}$ and $\vert y \vert < 2.5$ in an event.
This definition provides a general estimate of
the energy available to the parton-parton hard process and is less sensitive
to detector effects than ${\Pt}$. Major systematic uncertainties that hamper
jet cross section measurements, such as those due to the jet energy scale
and the jet selection efficiency, largely cancel in the measurement of $R_{32}$,
whilst the uncertainty on the integrated luminosity measurement cancels completely.
Hence, a measurement of $R_{32}$ is subject to smaller uncertainties
than the absolute multijet cross section
measurements and can provide a stringent test of the theoretical models.

Here the measured $R_{32}$ is used to perform a study on the validity of different
Monte Carlo (MC) models currently in use for modeling collisions
at the LHC.
The high centre-of-mass energy of the LHC allows a test of the models in a
transverse momentum region that was not accessible to previous experiments.
Such studies are important because QCD-induced processes constitute, in some cases,
the dominant backgrounds for signals of new physics at the LHC. This investigation
is complementary to other studies which also aim to probe the validity of pQCD early in the
LHC era~\cite{bib:EvSh-cms11, bib:AnDec-cms11, bib:AnDec-atlas11}.

The D0 Collaboration at the Fermilab Tevatron has measured the same ratio of cross
sections for jets in the range
$100 < H_{\rm T} < 600{\rm ~\GeV}$ and ${\rm p\overline{p}}$
centre-of-mass energy of $1.8\TeV$~\cite{bib:D0}.
Hence this study extends the range of the $R_{32}$ measurement
in a previously unexplored transverse momentum region,
and probes pQCD up to $ H_{\rm T} = 2.5 {\rm ~\TeV}$.

\section{The CMS Detector}

The central feature of the CMS apparatus is a superconducting solenoid
which provides an axial magnetic field of 3.8 T. The field volume of the
solenoid is instrumented with various particle detection systems.
Charged particle trajectories are measured by the silicon pixel and strip
trackers, with full azimuthal coverage within $\vert \eta \vert < 2.5$,
where the pseudorapidity $\eta$ is defined as $\eta = -\ln[\tan(\theta/2)]$,
and $\theta$ is the polar angle. The momentum resolution is about 1$\%$ at
100 \GeV. Surrounding the trackers are a lead-tungstate crystal electromagnetic
calorimeter (ECAL)~\cite{bib:ecal} with a preshower detector in the endcaps, which
covers the region $\vert \eta \vert < 3$, and a brass/scintillator hadron calorimeter
(HCAL)~\cite{bib:hcal}, which extends up to $ \vert \eta \vert = 5 $.
The ECAL has an energy resolution of better than 0.5\,\% for unconverted photons with
transverse energies above 100\,GeV.
The energy of charged pions and other quasi-stable hadrons can be measured with the
calorimeters (ECAL and HCAL combined) with a resolution of
$\Delta E/E \approx 100\,\% / \sqrt{E\,[\GeV]} \oplus 5\,\%$,
albeit with a nonlinear response.
For charged hadrons, the calorimeter resolution improves on the tracker momentum resolution
only for $\Pt$'s in excess of 500 $\GeV$.
The steel return yoke outside the solenoid is instrumented with gas-ionisation
detectors used to identify and reconstruct muons.
The CMS experiment collects data using a two-level trigger system: the first-level
hardware trigger (L1)~\cite{bib:trig-tdr} and the high-level software
trigger (HLT)~\cite{bib:hlt}. The L1 jet algorithm searches for collimated
transverse energy depositions using an effective sliding-window algorithm whose
size is 1.044 by 1.044 in ($\eta, \phi$) space.
Each jet candidate is assigned $\eta$, $\phi$, and $\Pt$ values that
are used to seed subsequent searches in the HLT. The HLT uses data at full
granularity and searches for jets using an iterative cone
algorithm~\cite{bib:iterative_cone, bib:cms-jet-alg} of
radius $R=\sqrt{(\Delta \eta) ^{2} + (\Delta\phi)^{2}} = 0.5$.
A more detailed description of the CMS detector can be found
in Ref.~\cite{bib:CMS_experiment}.

\section{Data Selection and Reconstruction}

The jet sample used for this study was collected using the L1 and HLT single jet
triggers. The efficiency of the trigger system is studied using a combination
of minimum-bias triggers~\cite{bib:cms-rap}, which are independent of the L1 jet
trigger, as well as low-threshold calorimeter triggers.  Five different jet-trigger
thresholds are used in this analysis, which ensure 100$\%$ trigger efficiency for
each of the different $H_{\rm T}$ regions of the jet sample
($\Ht >$ 0.20, 0.31, 0.40, 0.56, and 0.80 \TeV).
The integrated luminosities corresponding to each of the five regions are
$0.4$, $4.5$, $9.2$, $20.0$, and $36.0\ensuremath{\,\text{pb}^{-1}}$, respectively.

Each event is required to have an offline reconstructed vertex position along the beam line
that is within 24 cm of the nominal interaction point.
The global event reconstruction (also called particle-flow event reconstruction~\cite{bib:pf1})
consists in reconstructing and identifying each single particle with an optimised combination of
all sub-detector information. The energy of photons is directly obtained from the ECAL measurement,
corrected for zero-suppression effects. The energy of electrons is determined from a combination
of the track momentum at the main interaction vertex, the corresponding ECAL cluster energy,
and the energy sum of all bremsstrahlung photons attached to the track.
The energy of muons is obtained from the corresponding track momentum. The energy of charged hadrons
is determined from a combination of the track momentum and the corresponding ECAL and HCAL energy,
corrected for zero-suppression effects, and calibrated for the non-linear response of the calorimeters.
Finally the energy of neutral hadrons is obtained from the corresponding calibrated ECAL and HCAL energy.

For each event, hadronic jets are clustered from these reconstructed particles with the infrared
and collinear safe anti-$k_T$ algorithm, operated with a size parameter R of 0.5.
The jet momentum is determined as the vectorial sum of all particle momenta in this jet,
and is found in the simulation to be within 5$\%$ to 10$\%$ of the true momentum over the whole
$\Pt$ spectrum and detector acceptance.
An offset correction is applied to take into account the extra energy clustered in jets due
to additional proton-proton interactions within the same bunch crossing.
Jet energy corrections are derived from the simulation, and are confirmed with {\it in situ} measurements
with the energy balance of dijet and photon+jet events~\cite{bib:cms-jes}
Additional selection criteria are applied to each event to remove spurious jet-like
features originating from isolated noise patterns in certain HCAL regions.

Events with two or more jets of $H_{\rm T} > 0.2 {\rm ~\TeV}$, $\Pt > 50 {\rm ~\GeV}$,
and $\vert y \vert < 2.5$ are selected. The final dijet sample consists of $376\,590$ events.
A subsample of $206\,845$ events has an additional third jet that passes the $\Pt$ and $y$
selection criteria.

\section{Monte Carlo Models and Comparisons with Data}

The jet data are compared with the predictions of five QCD-based MC
generators that simulate jet production in pp collisions at $\sqrt{s}=7\,\mathrm{\TeV}$.

The {\sc Pythia} 6.4.22 MC generator~\cite{Sjostrand:2006za} uses LO matrix elements
to generate the $2 \rightarrow 2$ hard processes and a PS model to simulate higher-order processes.
The PS model gives a good description of parton emission when the emitted partons are
close in phase space. Events are generated with two different tunes of {\sc Pythia} 6.4:
D6T~\cite{Field:D6T} and Z2. The Z2 tune is identical to the Z1 tune described in~\cite{Field:D6T}
except that Z2 uses the CTEQ6L1~\cite{bib:cteq6l} proton parton distribution functions (PDF)
while Z1 uses CTEQ5L~\cite{bib:CTEQ5}.
Parton showers are ordered by mass in D6T and by $\Pt$ in Z2.
Hadronisation is simulated using the Lund string model~\cite{bib:lund1,bib:lund2}.

The {\sc Pythia} 8~\cite{Sjostrand:PYTHIA8} tune C2~\cite{PYTHIA8_C2} simulation orders
the PS in $\Pt$ and models the underlying event using the multiple parton interaction
model from {\sc Pythia}6, including initial- and final-state QCD radiation.

The {\sc Herwig++}~\cite{bib:Herwigpp} tune 2.3 program takes LO matrix elements and
simulates PS using the coherent-branching algorithm, with angular ordering of showers.
Hadronisation is simulated using the cluster model. The underlying event is generated
from the eikonal multiple parton-parton scattering model.

The {\sc MadGraph}~\cite{bib:MadGraph,bib:MadGraph2} generator is used because of its treatment
of hard and well-separated partons. {\sc MadGraph} generates tree-level helicity amplitudes
and is interfaced to {\sc Pythia}6, which generates the rest of the higher-order effects
using the PS model. Matching algorithms ensure that no double-counting occurs between the
tree-level and the PS-model-generated partons. The {\sc MadGraph} parameter that
distinguishes between PS and tree-level diagrams is set to 30 \GeV. The tune version is D6T.

The {\sc Alpgen}~\cite{bib:Alpgen} tune D6T MC program also generates tree-level processes.
As with {\sc MadGraph}, it is interfaced with {\sc Pythia}6 for generating higher-order effects.
The jet-parton matching threshold is set to 20 \GeV and the maximum distance between partons
and jets to $\Delta R = {\rm 0.7}$.

The CTEQ6L1 PDF set is used to model the proton structure in {\sc Pythia} 8 and {\sc MadGraph},
while {\sc Herwig++} and {\sc Alpgen} use the MRST2001~\cite{bib:MRST2001} and CTEQ5 PDF sets respectively.

The generated events from {\sc Pythia}6 tune Z2, {\sc MadGraph}, and {\sc Herwig++} are
processed through a full detector simulation based on {\sc Geant4}~\cite{bib:geant}.

Jets (denoted MC jets) are reconstructed from particle-flow objects by means of the anti-$k_T$ jet algorithm.
The jet energy corrections and selection criteria applied to MC events are identical to those applied to the data.
Jets are also reconstructed using the anti-$k_T$ jet algorithm on the MC four-vectors of the
stable particles after hadronisation. These (hadron-level) jets, referred to as GenJets,
represent the MC prediction free of detector effects.

Detailed studies of the $\Pt$, $y$, and $\Ht$ reconstruction are performed using the MC simulation.
The $\Pt$ resolution varies from 12\% at 50\GeV to 5\% at 1TeV.
These values are in agreement with the resolutions extracted from data.
The $\Ht$ resolution found using MC jets varies from 6$\%$ at 200 \GeV to 3.5$\%$ at 2.5 \TeV.

\begin{figure*}[htbp]
  \begin{center}
  \includegraphics[width=0.95\textwidth]{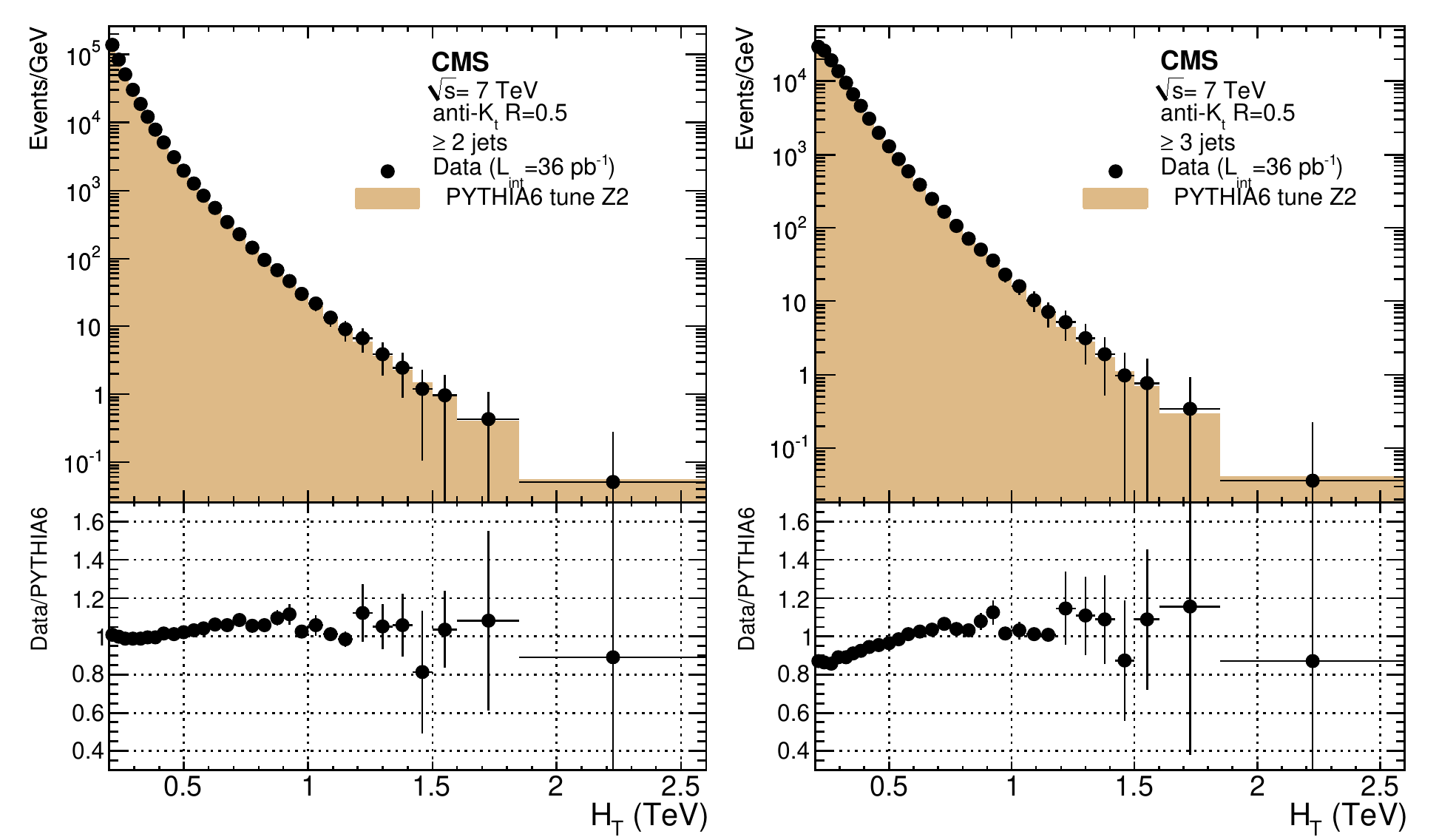}
  \caption{$\Ht$ distributions (top) for inclusive 2-jet (left) and 3-jet (right) events
           for data (solid circles) and {\sc Pythia}6 tune Z2 MC generator (histogram).
           Ratios between the MC predictions and the data (bottom).}
  \label{fig:ht_all}
\end{center}
\end{figure*}

Plotted in Fig.~\ref{fig:ht_all} (top) are the $\Ht$ distributions of the inclusive
2-jet (left) and 3-jet (right) samples after all selection criteria (solid circles).
The five different jet samples are normalised to the integrated luminosity
of $36 \ensuremath{\,\text{pb}^{-1}}$. The measurements are compared to the predictions
of {\sc Pythia}6 tune Z2 MC (histogram), normalised to the total number of dijet events.
Also shown in Fig.~\ref{fig:ht_all} (bottom) is the ratio between the MC predictions
and the data, separately for inclusive 2-jet and 3-jet events. The MC predictions describe
within $20\%$ the shape of the measured $\Ht$ distributions in data.This test has been
repeated using the predictions of {\sc MadGraph} and {\sc Herwig++} MC programs with
similar results.

\section{Extraction of \Ratio from the Data}

The inclusive 2-jet ($d\sigma_2/d\Ht$) and 3-jet ($d\sigma_3/d\Ht$) differential cross sections are
extracted from the data using

\begin{equation}
\label{eqn:inclusive_dijet_cross_section}
\frac{d\sigma_i}{d\Ht}=\frac{C_i}{L \varepsilon_i} \frac{N_i}{\Delta\Ht}~,
\end{equation}

where $C_i$, $\varepsilon_i$, and $N_i$ are the smearing correction, detection efficiency,
and the number of selected events for inclusive 2-jet ($i=2$) and inclusive 3-jet ($i=3$)
events respectively, $L$ is the integrated luminosity, and $\Delta\Ht$ is the bin size.

The smearing and efficiency corrections are computed using the {\sc Pythia}6 tune Z2:
GenJet events, consisting of all hadron-level events which have at least two jets with
$\ensuremath{{p}_\mathrm{T}} > 50\GeV$ and $\vert y \vert  < 2.5$, and
MC jet events, consisting of all events with at least two reconstructed jets
passing all selection criteria. The efficiency $\varepsilon_i$ is defined
as the number of selected events (at any reconstructed $\Ht$) arising from a given generated $\Ht$ interval
divided by the total number of generated events in that interval.
The efficiency for 2-jet events is $100\%$ throughout the $\Ht$ range, while the efficiency
for 3-jet events increases from $72\%$ at $\Ht=200$ \GeV to $100\%$
above $\Ht=400$ \GeV. The smearing correction $C_i$ is defined as the number of
selected events (at any reconstructed $\Ht$) arising from a given generated $\Ht$ interval divided by the
total number of selected events in that interval. The smearing correction $C_2$ is
approximately 0.9 throughout the $\Ht$ range, while $C_3$  is approximately 0.7
at $\Ht=200$ \GeV and increases to 0.9 for $\Ht>300$ \GeV.

The ratio \Ratio as a function of $\Ht$ is extracted from the data by dividing
the inclusive differential 3-jet cross section by the inclusive differential 2-jet
cross section:

\begin{equation}
\label{eqn:R32}
\Ratio = \frac{ d\sigma_{3}/{d\Ht} } { d\sigma_{2}/{d\Ht} } ~.
\end{equation}

The overall multiplicative correction factor to the data,
$C_3 /C_2 \times \varepsilon_2 / \varepsilon_3$, is within 4$\%$ of unity
in the range $0.2 < \ensuremath{\text{H}_\mathrm{T}} < 0.5\,\mathrm{\TeV}$
and within 2$\%$ thereafter. This method of extracting \Ratio is tested
by treating the {\sc MadGraph} sample as the data and using {\sc Pythia}6 as the MC.
The {\sc MadGraph} MC jets are treated and corrected (using {\sc Pythia}6) the same
way as the data and the extracted ratio from {\sc MadGraph} is compared with the ratio
at the hadron-level from {\sc MadGraph} using GenJets. The extracted ratio is within
$\pm 1\%$ of the GenJet ratio, and this difference is included in the overall
systematic uncertainty for $\Ratio.$

\begin{figure}[htbp]
  \begin{center}
  \includegraphics[width=0.80\columnwidth]{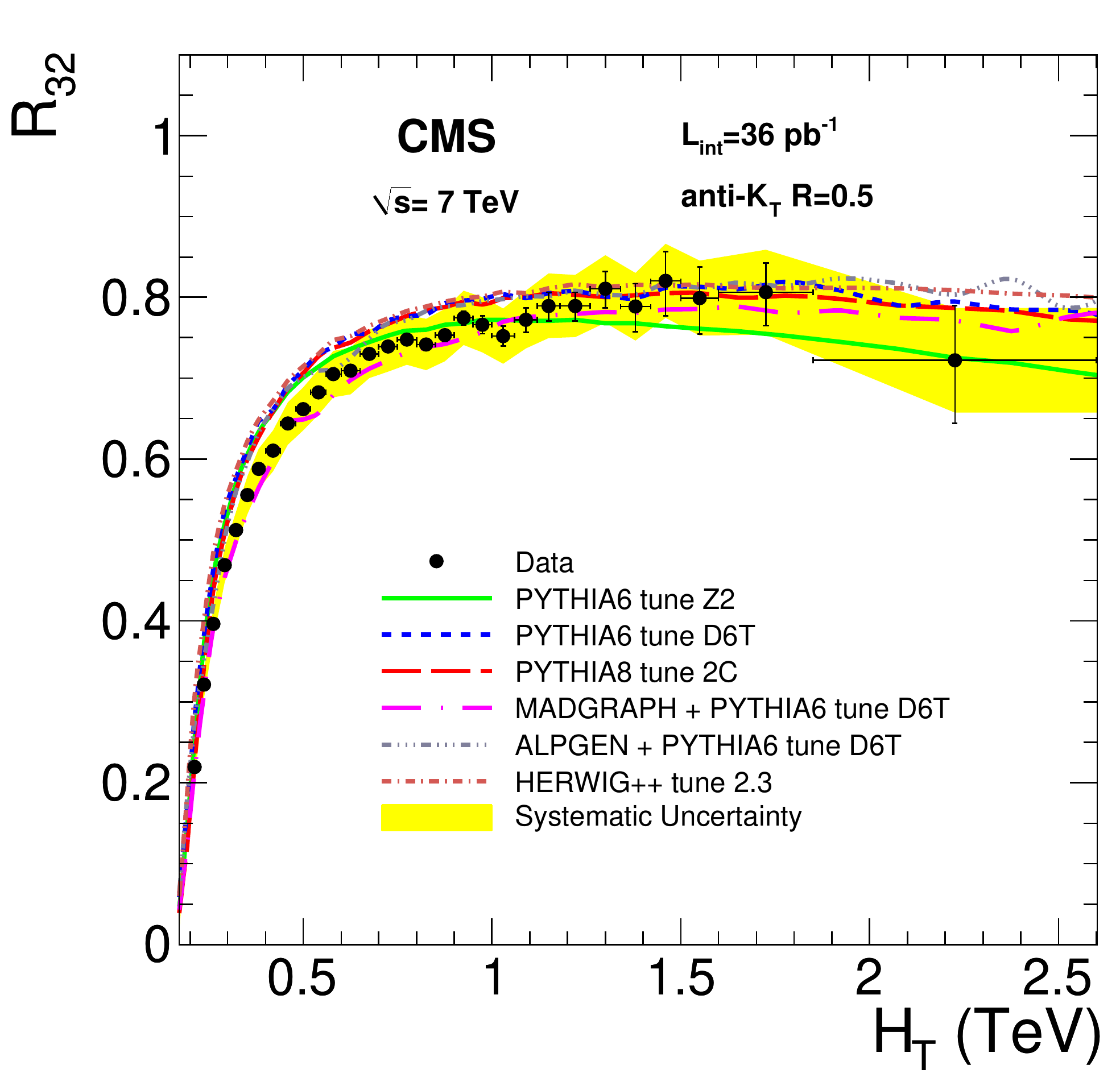}
  \caption{The measured \Ratio (solid circles) as a function of $\Ht$, and the predictions
           of {\sc Pythia}6, {\sc Pythia}8, {\sc MadGraph}, {\sc Alpgen}, and
		   {\sc Herwig++} (curves). Error bars represent statistical uncertainties. The
		   shaded area indicates the size of the combined systematic uncertainty. }
  \label{fig:R32_hadron_PYTHIA_mad}
  \end{center}
\end{figure}

The measured \Ratio after all corrections is shown in Fig.~\ref{fig:R32_hadron_PYTHIA_mad}.
As expected, \Ratio rises with increasing $\Ht$, as the phase space opens up for the emission
of a third jet satisfying the $\Pt>50\GeV$ selection criterion. The \Ratio value reaches
a plateau of about 0.8, which depends upon the $\Pt$ and $y$ selection criteria as well as the
jet-finding algorithm. The \Ratio plateau value measured here is different from that measured
by the D0 collaboration. The origin of this difference was studied using the MC and it is due
to the different centre-of-mass energy and $\Pt$ and $y$ selection criteria.

Several sources of systematic uncertainty on \Ratio are considered.
Uncertainties due to insufficient knowledge of the MC $\Ht$ distributions are estimated
by altering the shape of the MC distributions and comparing with the data.
Those uncertainties are in the range 3--8$\%$.
Uncertainties originating from the jet energy scale, the effect of pile-up,
insufficient knowledge of the jet $\Pt$ resolution, and self-consistency checks of the extraction method
(using {\sc MadGraph} as data) are considered.
The contribution of each of these effects is at the 1$\%$ level. The combined systematic uncertainty is
estimated by adding in quadrature all these uncertainties and is shown as the shaded area in
Fig.~\ref{fig:R32_hadron_PYTHIA_mad}. The combined systematic uncertainty is in the range 4--10$\%$.

\begin{figure*}[htbp]
  \begin{center}
  \includegraphics[width=0.95\textwidth]{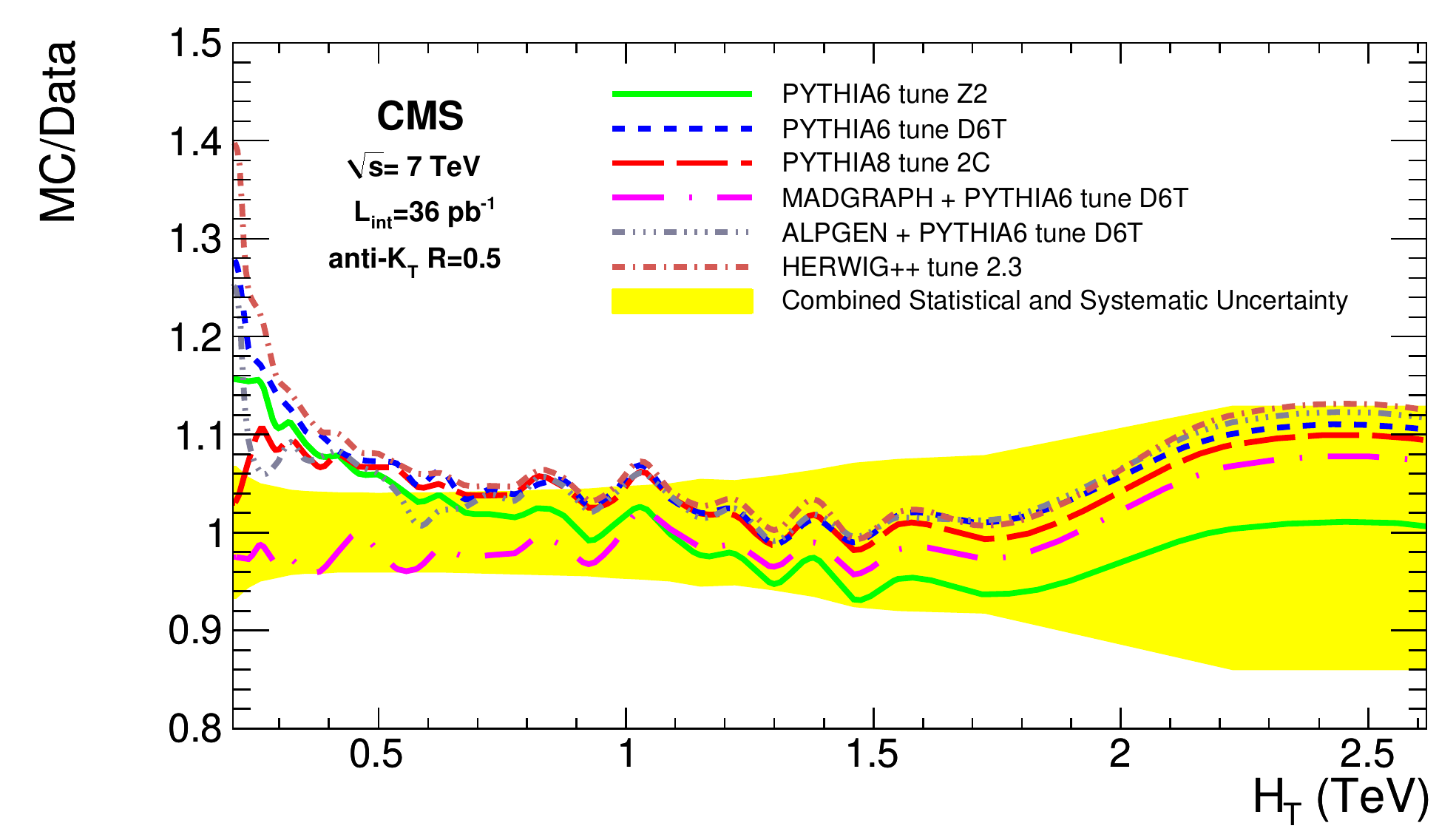}
  \caption{The ratios of the predicted \Ratio values from the {\sc Pythia}6, {\sc Pythia}8, {\sc MadGraph},
           {\sc Alpgen}, and {\sc Herwig++} Monte Carlo generators to the measured value, as a function
		   of $\Ht$. The shaded area indicates the size of the combined statistical and systematic uncertainty. }
  \label{fig:R32_MCoverData_total_latex}
  \end{center}
\end{figure*}

Also displayed in Fig.~\ref{fig:R32_hadron_PYTHIA_mad} are the predictions of the five MC generators
considered in this study. The ratios, \Ratio, of the different MC predictions to the measured
values are plotted in Fig.~\ref{fig:R32_MCoverData_total_latex}.
The shaded area represents the combined statistical and systematic uncertainty for the data.
The predictions of {\sc MadGraph} agree with the data throughout the $\Ht$ range of the measurement.
On the other hand, previous studies of event shapes and dijet angular
decorrelations~\cite{bib:EvSh-cms11, bib:AnDec-cms11}
indicate that {\sc MadGraph} does not describe these distributions well.
However, \Ratio is constructed from inclusive measurements and is mainly sensitive to the
probability of emitting a third parton in the final state, whereas the other two distributions
depend mainly on the final state of the event.
In contrast, the predictions of {\sc Alpgen}, {\sc Pythia}6, {\sc Pythia}8, and {\sc Herwig++}
are in agreement with the measured \Ratio for $\Ht > 0.5~{\rm \TeV}$,
although they overestimate it for lower values of $\Ht$.

\section{Summary}

The ratio of the inclusive 3-jet to 2-jet cross sections, for jets with
$\Pt > 50$ \GeV and $\vert y \vert < 2.5$, has been measured in the range
$0.2 < \Ht < 2.5 {\rm ~\TeV}$ for proton-proton collisions at a
centre-of-mass energy of $7\,\mathrm{\TeV}$. The measured ratio rises with
increasing $\Ht$, as the phase space opens for the production of a third jet,
reaching a plateau value of about 0.8 for $\Ht \gtrsim$ 1~\TeV.

This study tests the prediction of the different MC generators considered at \TeV scales.
{\sc Pythia}6 tune Z2, {\sc MadGraph}, and {\sc Herwig++}
describe within $20 \%$ the shape of the $\Ht$ distributions in data.

The predictions of {\sc MadGraph}, which generates tree-level helicity amplitudes,
are in agreement with the measured \Ratio throughout the range of this measurement.
{\sc Alpgen}, which also uses this method, describes well the data for $\Ht > 0.5 {\rm ~\TeV}$,
but overestimates \Ratio for lower $\Ht$ values.
The difference between the predictions of
{\sc MadGraph} and {\sc Alpgen} is an estimate of the uncertainty of the theoretical predictions
due to the different jet-parton matching parameters used by the two MC programs.
The current results combined with previous measurements of event shapes and dijet angular
decorrelations indicate that whilst {\sc MadGraph} has difficulties in describing the
details of the event shapes and dijet angular decorrelations, it predicts correctly the 
probability of emitting a third parton in the final state.
The MC programs that use LO matrix elements to simulate the hard
scattering and PS for all higher-order effects ({\sc Pythia}6, {\sc Pythia}8, {\sc Herwig++})
describe well the data above $\Ht> 0.5 {\rm ~\TeV}$, but overestimate \Ratio below this value.
This is the first measurement of this ratio in the multi-\!\TeV region, extending to an $\Ht$
range that has not been explored before.

\section*{Acknowledgements}

We wish to congratulate our colleagues in the CERN accelerator departments for the excellent
performance of the LHC machine. We thank the technical and administrative staff at CERN
and other CMS institutes, and acknowledge support from: FMSR (Austria); FNRS and FWO
(Belgium); CNPq, CAPES, FAPERJ, and FAPESP (Brazil); MES (Bulgaria); CERN; CAS, MoST,
and NSFC (China); COLCIENCIAS (Colombia); MSES (Croatia); RPF (Cyprus); Academy of
Sciences and NICPB (Estonia); Academy of Finland, ME, and HIP (Finland); CEA and CNRS/IN2P3 (France);
BMBF, DFG, and HGF (Germany); GSRT (Greece); OTKA and NKTH (Hungary);
DAE and DST (India); IPM (Iran); SFI (Ireland); INFN (Italy); NRF and WCU (Korea);
LAS (Lithuania); CINVESTAV, CONACYT, SEP, and UASLP-FAI (Mexico); PAEC (Pakistan);
SCSR (Poland); FCT (Portugal); JINR (Armenia, Belarus, Georgia, Ukraine, Uzbekistan);
MST and MAE (Russia); MSTD (Serbia); MICINN and CPAN (Spain); Swiss Funding Agencies
(Switzerland); NSC (Taipei); TUBITAK and TAEK (Turkey); STFC (United Kingdom); DOE and NSF (USA).

\clearpage

\clearpage

\bibliography{auto_generated}   
\appendix
\cleardoublepage \appendix\section{The CMS Collaboration \label{app:collab}}\begin{sloppypar}\hyphenpenalty=5000\widowpenalty=500\clubpenalty=5000\input{QCD-10-012-authorlist.tex}\end{sloppypar}
\end{document}

%% file: QCD-10-012-authorlist.tex
\textbf{Yerevan Physics Institute,  Yerevan,  Armenia}\\*[0pt]
S.~Chatrchyan, V.~Khachatryan, A.M.~Sirunyan, A.~Tumasyan
\vskip\cmsinstskip
\textbf{Institut f\"{u}r Hochenergiephysik der OeAW,  Wien,  Austria}\\*[0pt]
W.~Adam, T.~Bergauer, M.~Dragicevic, J.~Er\"{o}, C.~Fabjan, M.~Friedl, R.~Fr\"{u}hwirth, V.M.~Ghete, J.~Hammer\cmsAuthorMark{1}, S.~H\"{a}nsel, M.~Hoch, N.~H\"{o}rmann, J.~Hrubec, M.~Jeitler, G.~Kasieczka, W.~Kiesenhofer, M.~Krammer, D.~Liko, I.~Mikulec, M.~Pernicka, H.~Rohringer, R.~Sch\"{o}fbeck, J.~Strauss, F.~Teischinger, P.~Wagner, W.~Waltenberger, G.~Walzel, E.~Widl, C.-E.~Wulz
% ...
\vskip\cmsinstskip
% ...
1:~~Also at CERN, European Organization for Nuclear Research, Geneva, Switzerland\\
% ...
